# What are we weighting for?

A mechanistic model for probability weighting[*]

Ole Peters[†1,2], Alexander Adamou[‡1], Mark Kirstein[§1,3,4], and Yonatan Berman[¶1,5]

[1]London Mathematical Laboratory, London, UK
[2]Santa Fe Institute, NM, USA
[3]Max-Planck-Institute for Mathematics in the Sciences, Leipzig, Germany
[4]Institute of Mathematics, Leipzig University, Leipzig, Germany
[5]The Graduate Center and Stone Center on Socio-Economic Inequality, City University of New York, NY, USA

April 29, 2020

**Abstract**

Behavioural economics provides labels for patterns in human economic behaviour. Probability weighting is one such label. It expresses a mismatch between probabilities used in a formal model of a decision (*i.e.* model parameters) and probabilities inferred from real people's decisions (the same parameters estimated empirically). The inferred probabilities are called "decision weights." It is considered a robust experimental finding that decision weights are higher than probabilities for rare events, and (necessarily, through normalisation) lower than probabilities for common events. Typically this is presented as a cognitive bias, *i.e.* an error of judgement by the person. Here we point out that the same observation can be described differently: broadly speaking, probability weighting means that a decision maker has greater uncertainty about the world than the observer. We offer a plausible mechanism whereby such differences in uncertainty arise naturally: when a decision maker must estimate probabilities as frequencies in a time series while the observer knows them *a priori*. This suggests an alternative presentation of probability weighting as a principled response by a decision maker to uncertainties unaccounted for in an observer's model.

**JEL Codes**   C61 · D01 · D81

# 1   Introduction

*Probability weighting* is a concept that originated in prospect theory (KAHNEMAN and TVERSKY 1979; TVERSKY and KAHNEMAN 1992). It is one way to conceptualise a pattern in human behaviour of caution with respect to formal models. This is best explained by a thought experiment, in which

- a *disinterested observer* (DO), such as an experimenter, tells

- a *decision maker* (DM)

that an event occurs with some probability. The DO observes the DM's behaviour (*e.g.* gambling on the event) and finds it consistent with a behavioural model (*e.g.* expected-utility optimization) in which the DM uses a probability that differs systematically from what the DO has declared. The apparent probabilities, inferred from the DM's decisions, are called "*decision weights.*" We will adopt this nomenclature here.

- By "*probabilities,*" expressed as probability density functions (PDFs) and denoted $p(x)$, we will mean the numbers specified by a DO.

- By "*decision weights,*" also expressed as PDFs and denoted $w(x)$, we will mean the numbers that best describe the behaviour of a DM in the DO's behavioural model.[1]

Here, $x$ is a realisation of a random variable, $X$. For example, $X$ might be the payout of a gamble which the DM is invited to accept or decline.

Different behavioural models may result in different inferred decision weights. Our focus is not on how these weights are inferred, but on the robust observation that decision weights, $w(x)$ (used by DMs), are higher than probabilities, $p(x)$ (declared by DOs), for extreme events, *i.e.* when $p(x)$ is small. Thus, we do not consider any specific behavioural model: our aim is to find a general mechanistic explanation to probability weighting.

Probability weighting is often summarised visually by comparing

- cumulative density functions (CDFs) for probabilities, denoted

$$F_p(x) = \int_{-\infty}^{x} p(s)ds \ , \qquad (1)$$

- and CDFs for decision weights, denoted

$$F_w(x) = \int_{-\infty}^{x} w(s)ds \ . \qquad (2)$$

In Fig. 1 we reproduce the first such visual summary from TVERSKY and KAHNEMAN (1992, p. 310).

Plotting $F_w$ as a function of $F_p$ generates a curve, whose generic shape we shall call the "inverse-S". The inverse-S is the main observational finding in probability weighting: it sits above the diagonal $F_w = F_p$ for events of low cumulative probability (such that $F_w > F_p$ for these events) and below the diagonal for events of high cumulative probability (such that $F_w < F_p$).

---

[1] In the literature, decision weights are not always normalised, but for simplicity we will work with normalised decision weights. Mathematically speaking, they are therefore proper probabilities even though we don't call them that. Our results are unaffected because normalising just means dividing by a constant (the sum or integral of the non-normalised decision weights).

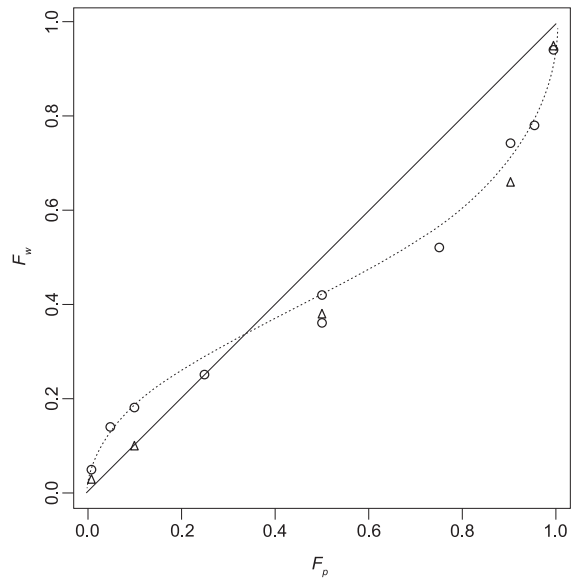

Figure 1: **Empirical phenomenon of probability weighting.** Cumulative decision weights $F_w$ (used by decision makers) versus cumulative probabilities $F_p$ (used by disinterested observers), as reported by TVERSKY and KAHNEMAN (1992, p. 310, Fig. 1, relabelled axes). The figure is to be read as follows: pick a point along the horizontal axis (the cumulative probability $F_p$ used by a DO) and look up the corresponding value on the vertical axis of the dotted inverse-S curve (the cumulative decision weight $F_w$ used by a DM). Low cumulative probabilities (left) are exceeded by their corresponding cumulative decision weights, and for high cumulative probabilities it's the other way around. It's the inverse-S shape of the curve that indicates this qualitative relationship.

As a final piece of nomenclature, we will use the terms *location*, *scale*, and *shape* when discussing probability distributions. Consider a standard normal distribution $\mathcal{N}(0,1)$, whose parameters indicate location 0 and squared scale 1. (For a Gaussian, the location is the mean and scale is the standard deviation.) For a general random variable $X$, with arbitrary location parameter $\mu_X$ and scale parameter $\sigma_X$, the following transformation produces a standardised random variable with identically-shaped distribution, but with location 0 and scale 1:

$$Z = \frac{X - \overbrace{\mu_X}^{\text{location}}}{\underbrace{\sigma_X}_{\text{scale}}} \ . \qquad (3)$$

Thus the PDF of $Z$, $p(z)$, is a density function with location $\mu_Z = 0$ and scale $\sigma_Z = 1$. In a graph of a distribution, a change of location shifts the curve to the left or right, and a change in scale shrinks or blows up the width of its features. Neither operation changes the *shape* of the distribution: two distributions have the same *shape* if they can be made to coincide through a linear transformation of the form (Eq. 3).

## 2 Probability weighting as a difference between models

Behavioural economics interprets Fig. 1 as evidence for a cognitive bias of the DM, an error of judgement. We will keep a neutral stance. We don't assume the DO to know "the truth" – he has a model of the world. Nor do we assume the DM to know "the truth" – he has another model of the world. From our perspective Fig. 1 merely shows that the two models differ. It says nothing about who is right or wrong.

### 2.1 The inverse-S curve

#### 2.1.1 Tversky and Kahneman

TVERSKY and KAHNEMAN (1992) chose to fit the empirical data in Fig. 1 with the following function,

$$\tilde{F}_w^{TK}(F_p; \gamma) = (F_p)^\gamma \frac{1}{[(F_p)^\gamma + (1-F_p)^\gamma]^{1/\gamma}} \, , \qquad (4)$$

which maps from one CDF, $F_p$, to another, $F_w$. We note that no mechanistic motivation was given for fitting this specific family of CDF mappings, parameterised by $\gamma$. The motivation is purely phenomenological: with $\gamma < 1$, this function can be made to fit to the data reasonably well.

The function $\tilde{F}_w^{TK}(F_p; \gamma)$ has one free parameter, $\gamma$. For $\gamma = 1$ it is the identity, and the CDFs coincide, $\tilde{F}_w^{TK}(F_p) = F_p$. Further, $\tilde{F}_w^{TK}$ has the following property: any curvature moves the intersection with the diagonal away from the mid-point $1/2$. This means if the function is used to fit an inverse-S (where $\gamma < 1$), the fitting procedure itself introduces a shift of the intersection to the left. We consider the key observation to be the inverse-S shape, whereas the shift to the left may be an artefact of the function chosen for the fit.

#### 2.1.2 Scale, location, and the inverse-S

We now make explicit how the robust qualitative observation of the inverse-S shape in Fig. 1 emerges when the DM uses a larger scale in his model of the world than the DO.

We illustrate this with a Gaussian distribution. Let's assume that a DO models an observable $x$ – which will often be a future change in wealth – as a Gaussian with location $\mu$ and variance $\sigma^2$. And let's further assume that a DM models the same observable as a Gaussian with the same location, $\mu$, but with a greater scale, so that the variance is $(\alpha\sigma)^2$. The DM simply assumes a broader range – $\alpha$ times greater – of plausible values, left panel of Fig. 2.

If the DM uses a greater scale in his model, then decision weights are higher than probabilities for low-probability events, and (because of normalisation) lower than probabilities for high-probability events. We can express this by plotting, for any value of $x$, the decision weight vs. the probability observed at $x$, right panel of Fig. 2.

In the Gaussian case we can write the distributions explicitly as

$$w(x) = \frac{1}{\sqrt{2\pi(\alpha\sigma)^2}} \exp\left[\frac{-(x-\mu)^2}{2(\alpha\sigma)^2}\right] \, , \qquad (5)$$

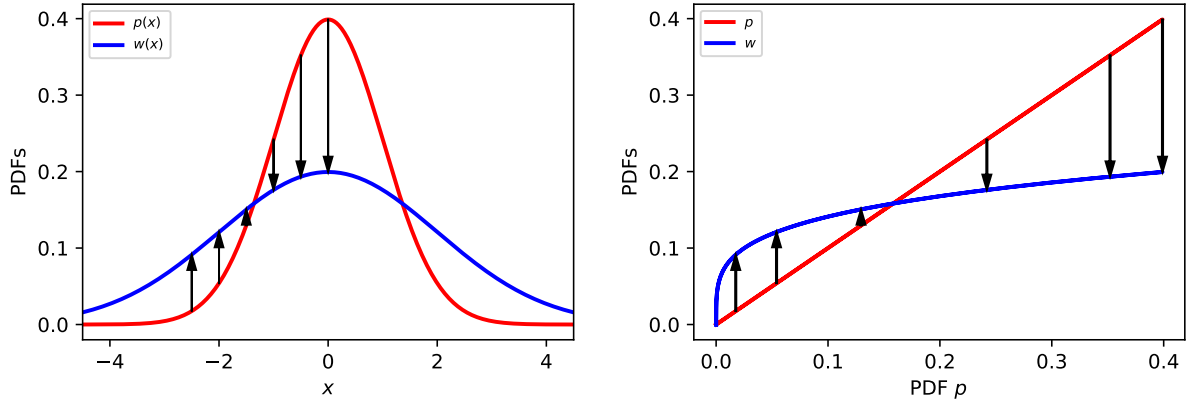

Figure 2: **Mapping PDFs.** Left: probability PDF (red), estimated by a DO; and decision-weight PDF (blue), estimated by a DM. The DO models $x$ with a best estimate for the scale (standard deviation) and assumes the true frequency distribution is the red line. The DM models $x$ with a greater scale (here 2 times greater, $\alpha = 2$), and assumes the true frequency distribution is the blue line. Comparing the two curves, the DM appears to the DO as someone who over-estimates probabilities of low-probability events and underestimates probabilities of high-probability events, indicated by vertical arrows. Right: the difference between decision weights and probabilities can also be expressed by directly plotting, for any value of $x$, the decision weight *vs.* the probability observed at $x$. This corresponds to a non-linear distortion of the horizontal axis. The arrows on the left correspond to the same $x$-values as on the right. They therefore start and end at identical vertical positions as on the left. Because of the non-linear distortion of the horizontal axis, they are shifted to different locations horizontally.

and
$$p(x) = \frac{1}{\sqrt{2\pi\sigma^2}} \exp\left[\frac{-(x-\mu)^2}{2\sigma^2}\right] \; . \qquad (6)$$

Eliminating $(x-\mu)^2$ from (Eq. 5) and (Eq. 6) yields the following expression for decision weight as a function of probability:

$$w(p) = p^{\frac{1}{\alpha^2}} \frac{(2\pi\sigma^2)^{\frac{1-\alpha^2}{2\alpha^2}}}{\alpha} \; . \qquad (7)$$

We plot this in the right panel of Fig. 2. As a sanity check, consider the shape of the $w(p)$ (blue curve, right panel Fig. 2): for a given value of $\alpha$, it is just a power law in $p$ with some pre-factor that ensures normalization. If $\alpha > 1$ it means that the DM uses a greater standard deviation than the DO. In this case, the exponent of $p$ satisfies $\frac{1}{\alpha^2} < 1$, and the blue curve is above the diagonal for small densities and below it for large densities.

Alternatively, we can express the difference between models by plotting the CDFs $F_w$ and $F_p$. We do this in Fig. 3, where the inverse-S emerges purely from the DM's greater assumed scale, $\alpha\sigma$.

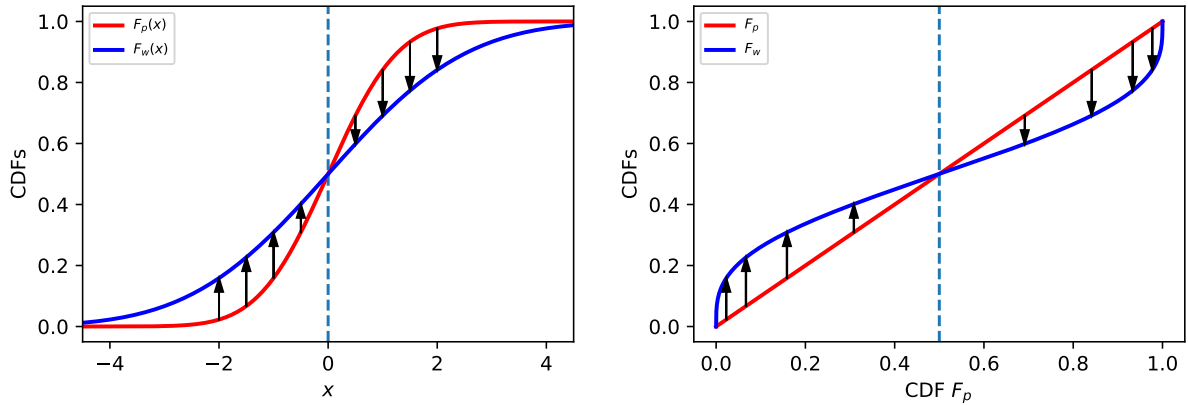

Figure 3: **Mapping CDFs.** Left: The DO assumes the observable $X$ follows Gaussian distribution $X \sim \mathcal{N}(0,1)$, which results in the red CDF of the standard normal, $F_p(x) = \Phi_{0,1}(x)$. The DM is more cautious, in his model the same observable $X$ follows a wider Gaussian distribution, $X \sim \mathcal{N}(0,4)$ depicted by $F_w(x)$ (blue). Following the vertical arrows (left to right), we see that for low values of the event probability $x$ the DM's CDF is larger than the DO's CDF, $F_p(x) < F_w(x)$; the curves coincide at 0.5 because no difference in location is assumed; necessarily for large values of the event probability $x$ the DM's CDF must be lower than the DO's. Right: the same CDFs as on the left but now plotted not against $x$ but against the CDF $F_p$. Trivially, the CDF $F_p$ plotted against itself is the diagonal; the CDF $F_w$ now displays the generic inverse-S shape known from prospect theory. The arrows start and end at the same vertical values as on the left.

## 2.2  Different scales and locations

In Fig. 4 we explore what happens if both the scales and the locations of the DO's and DM's models differ. Visually, this produces an excellent fit to empirical data, to which we will return in Sec. 4. A difference in assumed scales and locations, for simple Gaussian distributions, is sufficient to reproduce the observations. This suggests a different nomenclature and a conceptual clarification. The inverse-S curve does not mean that "probabilities are re-weighted." It means only that experimenters and their subjects have different views about appropriate models of, and responses to, a situation.

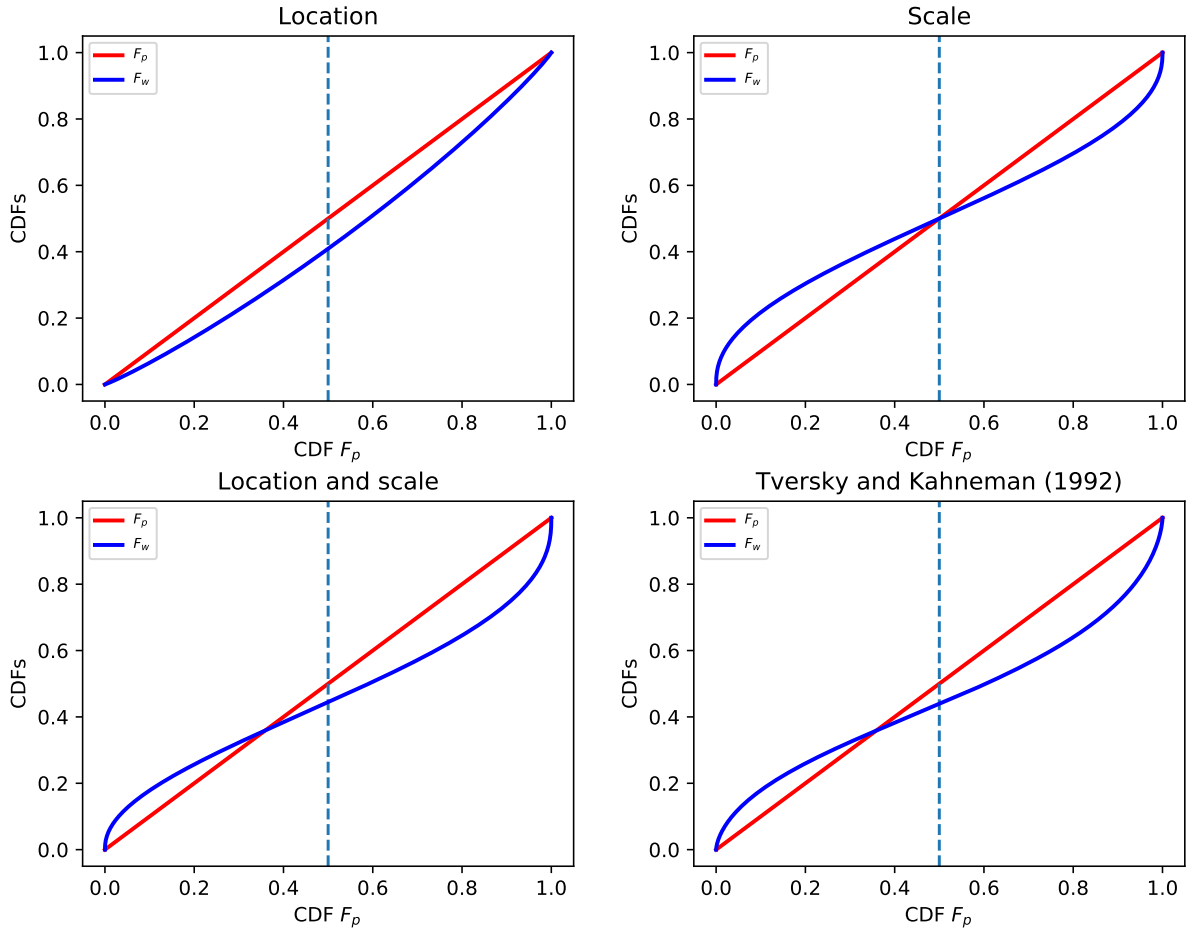

Figure 4: **CDF maps for Gaussian distributions.** Top left: Difference in location. DO assumes location 0, scale 1; DM assumes location 0.23 (bigger than DO), scale 1. Top right: Difference in scale. DO assumes location 0, scale 1; DM assumes location 0, scale 1.64 (broader than DO). Bottom left: Differences in scale and location. DO assumes location 0, scale 1; DM assumes location 0.23 (bigger than DO), scale 1.64 (broader than DO). Bottom right: Fit to observations reported by TVERSKY and KAHNEMAN (1992). This is (Eq. 4) with $\gamma = 0.65$. Note the similarity to bottom left.

## 2.3 Different shapes

Numerically, our procedure can be applied to arbitrary distributions:

1. construct a list of values for the CDF assumed by the DO, $F_p(x)$.

2. construct a list of values for the CDF assumed by the DM, $F_w(x)$.

3. plot $F_w(x)$ vs. $F_p(x)$.

Of course, the DM could even assume a distribution whose shape differs from that of the DO's distribution. The inverse-S arises whenever a DM assumes a greater scale for a unimodal distribution. To illustrate the generality of the procedure, in Fig. 5 we carry it out for Student's (power-law tailed) $t$-distributions (which we refer to as $t$-distributions), where DO and DM use different shape parameters and different locations [2] The result is qualitatively similar to the bottom right panel of Fig. 3, corresponding to (Eq. 4).

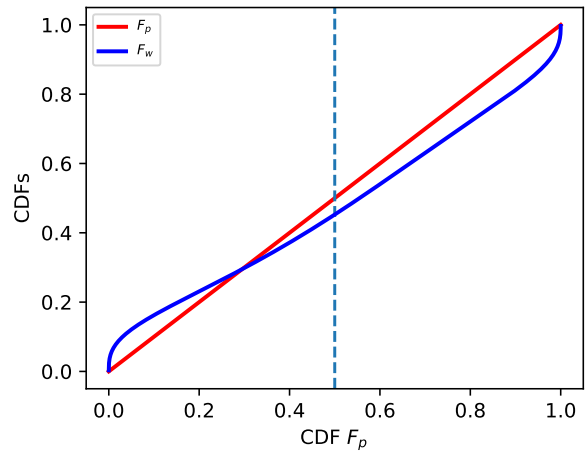

Figure 5: **Probability weighting for $t$-distributions.** The DM uses different shape and location parameters (1 and 0.35 respectively) from those of the DO (3 and 0.2).

## 3   Reasons for different models

Probability weighting is usually interpreted as a cognitive bias that leads to errors of judgement and poor decisions by DMs.[3] We caution against this interpretation. At least we should keep in mind that it is unclear who suffers from the bias: experimenter or test subject (or neither or both)? We are by no means the first to raise this question. Commenting on another so-called cognitive bias regarding probabilities, the representativeness fallacy, COHEN (1979) asked: "Whose is the fallacy?"

---

[2]The PDF of the $t$-distribution is

$$p(x) = \frac{\Gamma\left(\frac{\nu+1}{2}\right)}{\Gamma\left(\frac{\nu}{2}\right)\sqrt{\pi\nu}\sigma}\left(1+\frac{1}{\nu}\left(\frac{x-\mu}{\sigma}\right)^2\right)^{-\frac{\nu+1}{2}}, \qquad (8)$$

where $\nu$ is the shape parameter, $\sigma$ is the scale parameter, and $\mu$ is the location parameter. The corresponding CDF is

$$F(x) = \begin{cases} 1 - \frac{1}{2}I_{\frac{\nu}{\left(\frac{x-\mu}{\sigma}\right)^2+\nu}}\left(\frac{\nu}{2}, \frac{1}{2}\right) & \text{if } x-\mu \geq 0; \\ \frac{1}{2}I_{\frac{\nu}{\left(\frac{x-\mu}{\sigma}\right)^2+\nu}}\left(\frac{\nu}{2}, \frac{1}{2}\right) & \text{if } x-\mu < 0, \end{cases} \qquad (9)$$

where $I_x(a,b)$ is the incomplete beta function.

In the limit $\nu \to \infty$, the $t$-distribution converges to a Gaussian with location $\mu$ and scale $\sigma$. We assume by default that $\sigma = 1$, so the $t$-distribution is effectively characterised by two parameters: shape ($\nu$) and location ($\mu$).

[3]Indeed, its originators presented it as such. Introducing prospect theory, KAHNEMAN and TVERSKY (1979, p. 277) wrote "we are compelled to assume [...] that decision weights do not coincide with stated probabilities. These departures from expected utility theory must lead to normatively unacceptable consequences". They classified prospect theory as descriptive rather than normative, *i.e.* as relating to actual rather than optimal behaviour (TVERSKY and KAHNEMAN 1986, p. S252). Put simply, prospect theory aims to model systematic errors in human decision making, arising (in part) from inappropriate psychological adjustments of known probabilities.

Whatever the answer, two observations are robust and interesting: first, disagreement is common; and, second, the disagreement tends to go in the same direction, with DMs assuming a greater range of plausible outcomes than DOs.

An explanation for the first observation is that probability is a slippery concept and the word is used to mean different things. This suggests that phrasing information about probabilities concretely should reduce disagreement between DO and DM. For example, the statement "10 out of 100 people have this disease" conveys more, and more precise, information than "the probability of having this disease is 0.1." Specifically, it tells us that a sample of people has been observed and what the size of the sample is.

Furthermore, GIGERENZER (2018) argues that statements involving integer counts, or what he calls natural frequencies, ("10 out of 100") are more readily understood by people than statements involving fractional probabilities ("0.1").

The second observation may be explained as follows. A DO often has control over, and essentially perfect knowledge of, the decision problem he poses. A DM does not have such knowledge, and this ignorance will often translate into additional assumed uncertainty. For example, the DO may know the true probabilities of some gamble in an experiment, while the DM may have doubts about the DO's sincerity and his own understanding of the rules of the game. We will return to this in Sec. 3.2.

## 3.1 Some meanings of "probability"

Many thousands of pages have been written about the meaning of probability. We will not attempt a summary of the philosophical debate and instead highlight a few relevant points.

**Frequency-in-an-ensemble interpretation of probability**

Consider the simple probabilistic statement: "the probability of rain here tomorrow is 70%." Tomorrow only happens once, so one might ask: in 70% of what will it rain? The technical answer to this question is often: rain happens in 70% of the members of an ensemble of computer simulations, run by a weather service, of what may happen tomorrow. So one interpretation of "probability" is "relative frequency in a hypothetical ensemble of simulated possible futures."

It is thus a statement about a model. How exactly it is linked to physical reality is not completely clear.

**Frequency-over-time interpretation of probability**

In some situations, the statement "70% probability of rain here tomorrow" refers to the relative frequency over time. Before the advent of computer models in weather forecasting, people used to compare today's measurements (of, say, wind and pressure) to those from the past – weeks, months, or even years earlier. Forecasts were made on the assumption that the weather tomorrow would resemble the weather that had followed similar conditions in the historical record.

Rather than a statement about outcomes of an *in silico* model, the statement may thus be a summary of real-world observations over a long time.

**Degree-of-belief interpretation of probability**

No matter how "probability" relates to a frequentist physical statement, whether with respect to an ensemble of simultaneously possible futures or to a sequence of actual past futures, it also corresponds to a mental state of believing something with a degree of conviction: "I'm 90% sure I left my wallet in that taxi."

For our purpose it suffices to say that there's no guarantee that a probabilistic statement will be interpreted by the receiver (the DM) as it was intended by whoever made the statement (the DO).

## 3.2 Consistent differences between DO and DM

**Estimation errors for probabilities**

Let's assume that both the DO and the DM mean by "probability" the relative frequency of an event in an infinitely long time series of observations. Of course, real time series have finite length, so probabilities defined this way are model parameters and cannot actually be observed. But, from a real time series, we can estimate the best values to put into a model, by counting how often we see an event.

As the probability of an event gets smaller, so does the number of times we see it in a finite time series. If we want to say something about the uncertainty in this number, we can measure it – or imagine measuring it – in several time series to see how much it varies. The variations from one time series to another get smaller for rarer events, but the *relative* variations get larger, and so does the relative uncertainty in our estimate of probabilities. Take an extreme simplified example: asymptotically an event occurs in 0.1% of observations, and we have a time series of 100 observations. Around 99.5% of such time series will contain 0 or 1 events. Naïvely, then, we would estimate the probability as either 0 or 1%. In other words, we would estimate the event as either impossible or occurring ten times more frequently than it really would in a long series. However, if the event occurs 50% of the time asymptotically, then around 99.5% of time series would contain between 35 and 65 events, leading to a much smaller relative error in probability estimates.

A DM who must estimate probabilities from observations is well advised to account for this behaviour of uncertainties in his decision making. Specifically, the DM should acknowledge that, due to his lack of information, *prima facie* rare events may be rather more common than his data suggest, while common events, being revealed more often, are more easily characterised. In such circumstances, caution may dictate that the DM assign to rare events higher probabilities than his estimates, commensurate with his uncertainty in them. This would look like probability weighting to a DO and, indeed, would constitute a mechanistic reason for it.[4]

Formalising these thoughts, we find that so long as relative uncertainties are larger for rare events than for common events – which, generically, they are – then an inverse-S curve emerges. See Appendix A for a detailed discussion. Here we make a simple scaling argument and then check it with a simulation. For an asymptotic probability density $p(x)$, the number of events $n(x)$ we see in the small interval $[x, x+\delta x]$ in a time series of $T$ observations is proportional to $p(x)$, to $\delta x$, and to $T$. So we have $n(x) \sim p(x)\delta x T$, where we mean by $\sim$ "scales like." We also know that such counts, for example in the simple Poissonian case, are random variables whose uncertainties scale like $\sqrt{n(x)}$.

---

[4]Interestingly, KAHNEMAN and TVERSKY (1979, p. 281) made the same point, noting that "overestimation that is commonly found in the assessment of the probability of rare events" has the same effect on human decisions as probability weighting. Since they assumed that subjects in experiments adopt unquestioningly the stated probabilities, they argued that probability weighting was necessary to explain their observations. We make no such assumption here.

If we knew the asymptotic probability density $p(x)$, we could make an estimate of the count as

$$n(x) \approx p(x)\delta x T \pm \sqrt{p(x)\delta x T} \ . \tag{10}$$

We would write $\hat{n}(x) \equiv p(x)\delta x T$ as the estimate of $n(x)$ and $\varepsilon\left[\hat{n}(x)\right] \equiv \sqrt{p(x)\delta x T}$ as its uncertainty. Of course, this situation seldom applies, because usually we do not know $p(x)$.

Conversely, and more realistically, if we observe a count $n(x)$, then we can use the scaling $p(x) \sim n(x)/T\delta x$ to make an estimate of the asymptotic probability density as

$$p(x) \approx \frac{n(x)}{T\delta x} \pm \frac{\sqrt{n(x)}}{T\delta x} \ . \tag{11}$$

We write $\hat{p}(x) \equiv n(x)/T\delta x$ as the estimate of $p(x)$, and

$$\varepsilon\left[\hat{p}(x)\right] \equiv \frac{\sqrt{n(x)}}{T\delta x} = \sqrt{\frac{\hat{p}(x)}{T\delta x}} \tag{12}$$

as its uncertainty, which we have expressed in terms of the estimate itself.

The standard error, $\sqrt{\hat{p}(x)/T\delta x}$, in an estimated probability density shrinks as the probability decreases. However, the relative error in the estimate is $1/\sqrt{\hat{p}(x)T\delta x}$, which grows as the event becomes rarer. This is consistent with our claim, that low probabilities come with larger relative errors, and constitutes the key message of this section. Errors in probability estimates behave differently for low probabilities than for high probabilities: absolute errors are smaller for lower probabilities, but relative errors are larger.

Let's assume that the DM is aware of the uncertainties in his estimates and, furthermore, that he does not like surprises. To avoid surprises, he adds the standard error to his estimate of the probability density, $\hat{p}(x)$, in order to construct his decision weight density, $w(x)$. In effect, he constructs a reasonable worst case for each of his estimates. After normalising, this conservative strategy yields generically,

$$w(x) = \frac{\hat{p}(x) + \varepsilon\left[\hat{p}(x)\right]}{\int_{-\infty}^{\infty} \left(\hat{p}(s) + \varepsilon\left[\hat{p}(s)\right]\right) ds} \ , \tag{13}$$

and specifically, for the type of uncertainty we consider,

$$w(x) = \frac{\hat{p}(x) + \sqrt{\frac{\hat{p}(x)}{T\delta x}}}{\int_{-\infty}^{\infty} \left(\hat{p}(s) + \sqrt{\frac{\hat{p}(s)}{T\delta x}}\right) ds} \ . \tag{14}$$

Note that the cautionary correction term in (Eq. 14) is parametrised by $T\delta x$, which scales like the number of observations in $[x, x+\delta x]$. As $T\delta x$ grows large, the correction vanishes and both $w(x)$ and $\hat{p}(x)$ become consistent with the asymptotic density, $p(x)$. With perfect information, a DM need not adjust decisions to account for uncertainty.

Does our analysis, culminating in (Eq. 13) and (Eq. 14), reproduce the stylised facts of probability weighting, in particular the inverse-S curve? We check in two ways. First, analytically, by applying the DM's cautionary correction in (Eq. 14) directly to reference probability density functions. Second, by simulating the DM compiling counts of outcomes drawn from reference distributions, from which he estimates probability densities and their uncertainties. The simulation is meant to explore how noisy the effect is when a DM really only sees a single time series. The Python code is available at `bit.ly/lml-pw-code-dm-count`, and a Jupyter notebook can be loaded to manipulate the code in an online environment at `bit.ly/lml-pw-dm-count-b`. In both cases, we treat the DO as using the reference distribution to make his predictions of the DM's behaviour.

Figure 6 shows the resulting PDFs and CDF mappings generated by setting $\hat{p}(x)$ in (Eq. 14) to be the probability density functions for a Gaussian distribution and a fat-tailed $t$-distribution. Inverse-S curves are found for both distributions and the effect is more pronounced for the fat-tailed distribution.

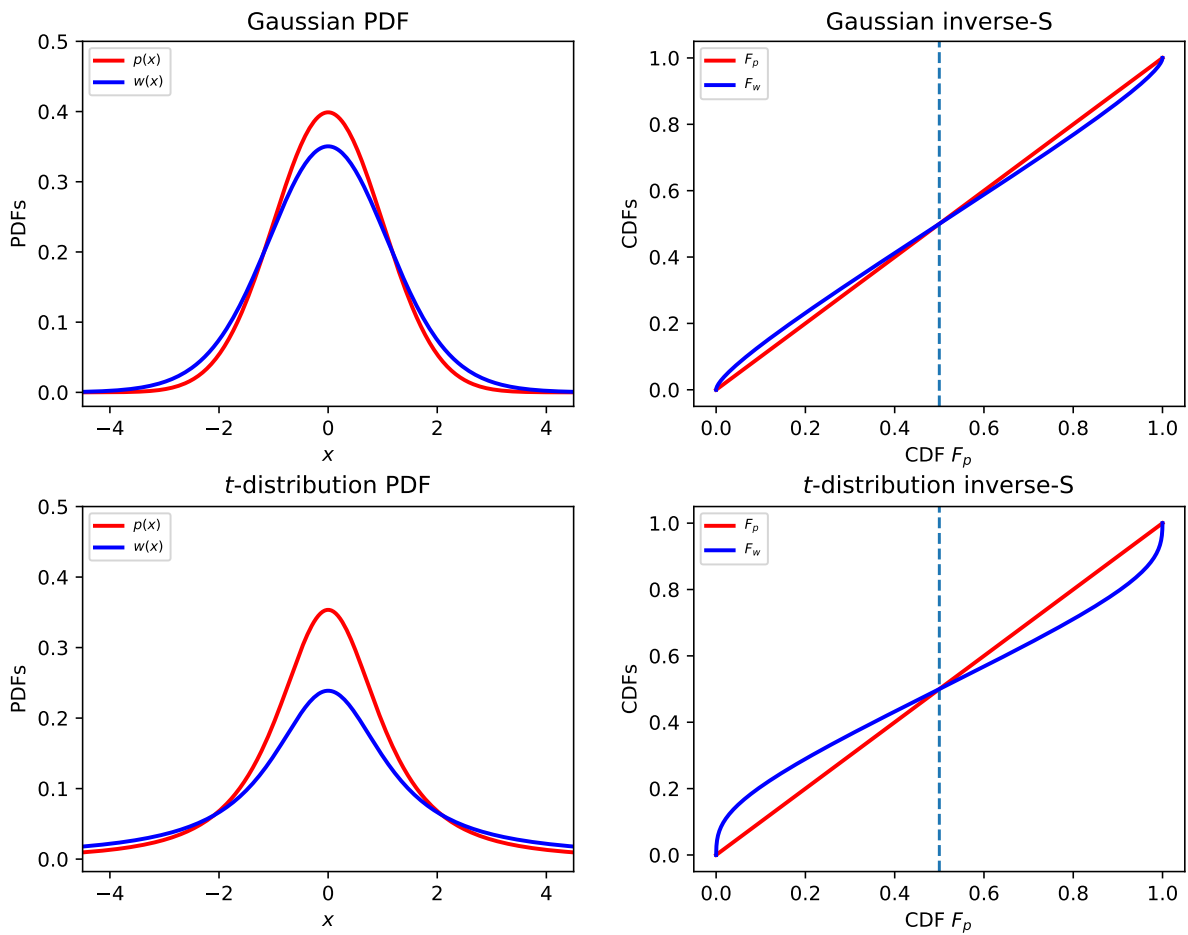

Figure 6: **Mapping PDFs and CDFs with estimation errors.** PDFs (left) and inverse-S curves (right) arising when the DO assumes a Gaussian (scale 1, location 0, top line) or a $t$-distribution (shape 2, location 0, bottom line), and the DM uses decision weights according to (Eq. 14) with $T\delta x = 10$. For the fat-tailed $t$-distribution (in the bottom line) the difference between $p(x)$ and $w(x)$ is more pronounced.

Figure 7 shows the results of a computer simulation of a DM who observes a series of realisations of either Gaussian or $t$-distributed random variables, which he counts into bins. In the simulation, a probability density, $\hat{p}(x)$, is estimated for each bin as $n(x)/T\delta x$ and its uncertainty, $\varepsilon\left[\hat{p}(x)\right]$, is obtained numerically as standard deviation in each $\hat{p}(x)$ over 1000 parallel simulations. The DM's decision weights are then obtained according to (Eq. 13). Again, inverse-S curves are found for both distributions, corroborating our scaling arguments.

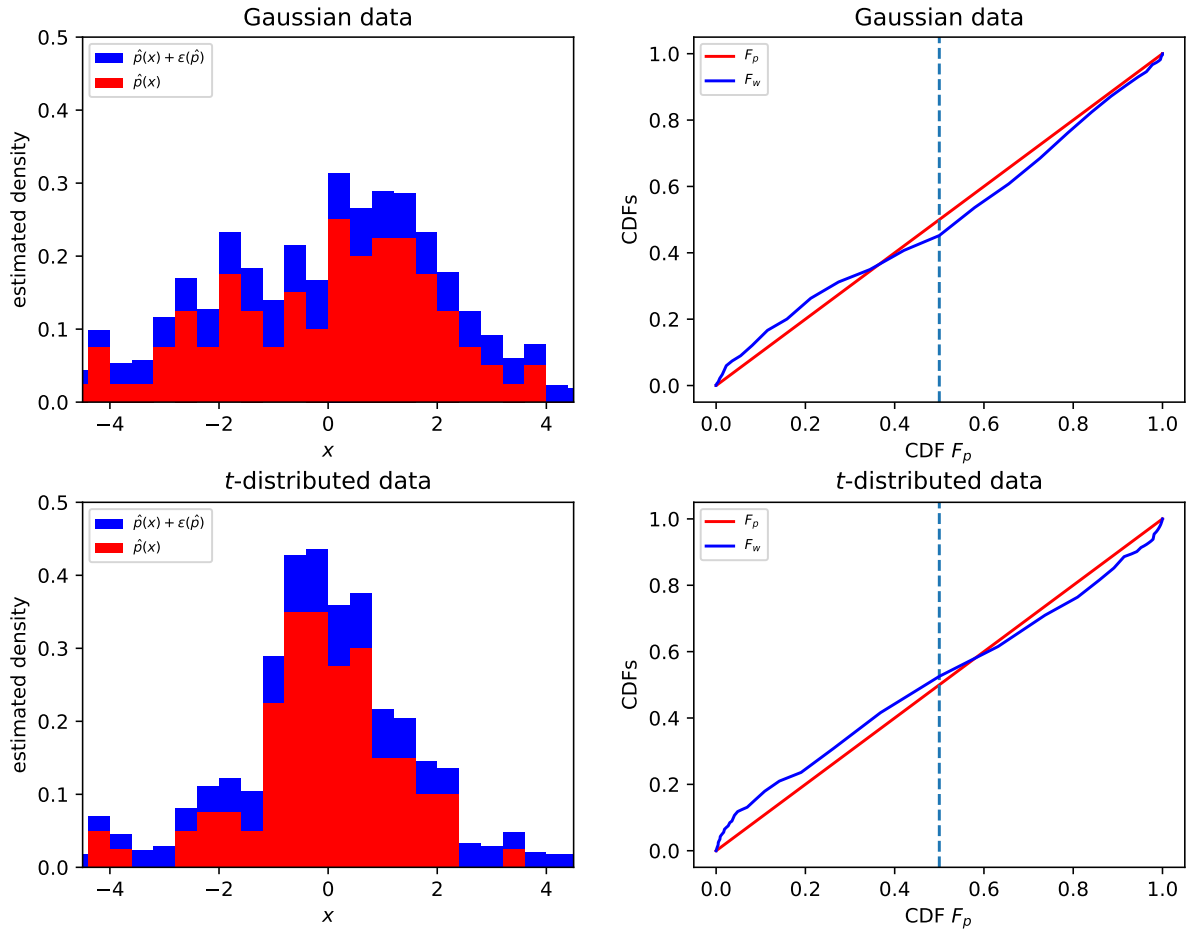

Figure 7: **Simulations of a DM estimating probability densities by counting events in a finite series of observations.** Left: estimated probability densities for $T = 100$ Gaussian (top; location 0, scale 2) and $t$-distributed (bottom; location 0, scale 1, shape 1.5) variates counted in bins of width $\delta x = 0.4$. Red bars show the estimates, $\hat{p}(x)$, and blue bars show the estimates with one standard error added, $\hat{p}(x) + \varepsilon\left[\hat{p}(x)\right]$. Right: inverse-S curves for a DO who assumes $p(x)$ follows a Gaussian (top) and $t$-distribution (bottom), while the DM uses decision weight density, $w(x)$, derived by normalising his conservative estimates (blue bars on left) according to (Eq. 13).

### 3.2.1 Typical situations of DO and DM: ergodicity

To recap: behavioural economists observe that DOs tend to assign lower weights to low-probability events than DMs. While behavioural economists commonly assume that the DM is wrong, we make no such judgement. In any decision problem, the aim of the decision must be taken into account. Crucially, this aim depends on the situation of the individual.

The two types of modellers (DO and DM) pursue different goals. In our thought experiment, the DO is a behavioural scientist without personal exposure to the success or failure of the DM, whom we imagine as a test subject or someone whose behaviour is being observed in the wild. The DM, of course, has such exposure. Throughout the history of economics, it has been a common mistake, by DOs, to assume that DMs optimise what happens to them on average in an ensemble. To the DM, what happens to the ensemble is seldom a primary concern. Instead, he is concerned with what happens to him over time. Not distinguishing between these two perspectives is only permissible if they lead to identical predictions, meaning only if the relevant observables are ergodic (PETERS 2019).

It is now well known that this is usually not the case in the following sense: DMs are usually observed making choices that affect their wealth, and wealth is usually modelled as a stochastic process that is not ergodic. The ensemble average of wealth does not behave like the time average of wealth.

The most striking example is the universally important case of noisy multiplicative growth, the simplest model of which is geometric Brownian motion, $dx = x(\mu dt + \sigma dW)$. In the present context of human economic decisions, this is the most widely used model of the evolution of invested wealth. The average over the full statistical ensemble (often studied by the DO) of geometric Brownian motion grows as $\exp(\mu t)$. Each individual trajectory, on the other hand, grows in the long run as $\exp[(\mu - \frac{\sigma^2}{2})t]$. If the DO takes the ensemble perspective, he will deem the fluctuations irrelevant whereas, from the DM's time perspective, they reduce growth. So, while a DO curious about the ensemble may suffer no consequences from disregarding rare events, hedging against such events is central to the DM's success.

The difference between how these two perspectives evaluate the effects of probabilistic events is qualitatively in line with the observed phenomena we set out to explain. The DM typically has large uncertainties, especially for low-probability events, and has an evolutionary incentive to err on the side of caution, *i.e.* to behave as though extreme events have a higher probability than in the DO's model.

## 4 Fitting the model to experimental results

Visually, looking at the figures and the level of noise in the data in Fig. 1, one would conclude that Tversky and Kahneman's physically unmotivated function, $\tilde{F}_w^{TK}(F_p)$ in (Eq. 4), fits the data no more efficiently than the functions arising from our mechanistic model. This is particularly evident in the bottom panels of Fig. 4, which show that a Gaussian, $w(x)$, whose scale and location differ from those of $p(x)$, reproduces the fitted functional shape of $\tilde{F}_w^{TK}(F_p)$.

For completeness and scientific hygiene, in the present section we fit location and scale parameters in the Gaussian and $t$ models for $F_w$ to experimental data from TVERSKY and KAHNEMAN (1992) (depicted in circles in Fig. 1) and from TVERSKY and FOX (1995). Specifically, in the Gaussian model we fit the location and scale parameters $\mu$ and $\sigma$ in the CDF,

$$F_w(x) = \Phi\left(\frac{\Phi^{-1}(F_p(x)) - \mu}{\sigma}\right), \qquad (15)$$

where $\Phi$ is the CDF of the standard normal distribution. In the $t$-model, we fit the location and shape parameters, $\mu$ and $\nu$, in the CDF, $F_w(x)$, of a $t$-distributed random variable (see Sec. 2.3). In both cases, we assume that $F_p(x)$ is that of a standard normal distribution.

In addition to (Eq. 4) used by Tversky and Kahneman, we fit the function

$$\tilde{F}_w^L(F_p; \delta, \gamma) = \frac{\delta F_p^\gamma}{\delta F_p^\gamma + (1-F_p)^\gamma}, \qquad (16)$$

suggested by LATTIMORE et al. (1992) to parametrically describe probability weighting (also used by TVERSKY and WAKKER (1995) and PRELEC (1998)). The reason for fitting (Eq. 16) is to ensure a fair comparison: the Gaussian and $t$ models are characterised by two parameters, whereas (Eq. 4) only has one free parameter. Equation (16) has two parameters.

Figure 8 presents the fit results. We obtain very good fits to data for both Gaussian and $t$-distributions, as well as for (Eq. 4) and (Eq. 16), in the two experiments. It is practically impossible to distinguish between the fitted functions within standard errors. We conclude that our model fits the data well, and unlike (Eq. 4) or (Eq. 16), the fitted functions are directly derived from a physically plausible mechanism. They are not simply phenomenological.

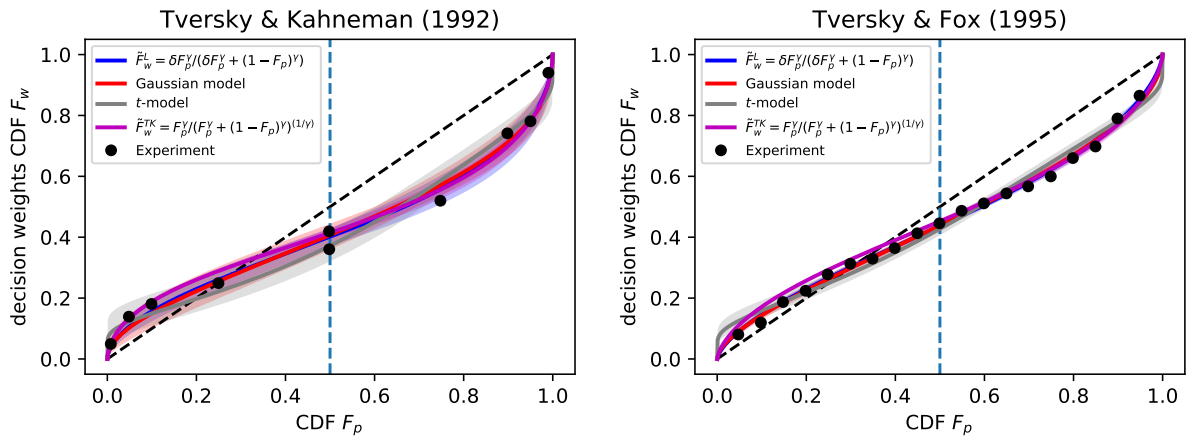

Figure 8: **Model fitting to experimental data from Tversky and Kahneman (1992) (left) and Tversky and Fox (1995) (right).** Left: LATTIMORE et al. (1992) (Eq. 16): $\delta = 0.67$ ($SE = 0.04$), $\gamma = 0.58$ ($\pm 0.03$); Gaussian model: $\mu = 0.38$ ($\pm 0.06$), $\sigma = 1.60$ ($\pm 0.10$); $t$-model: $\nu = 1.27$ ($\pm 0.28$), $\mu = 0.40$ ($\pm 0.07$); TVERSKY and KAHNEMAN (1992) (Eq. 4): $\gamma = 0.60$ ($\pm 0.02$). Right: LATTIMORE et al. (1992): $\delta = 0.77$ ($\pm 0.01$), $\gamma = 0.69$ ($\pm 0.01$); Gaussian model: $\mu = 0.22$ ($\pm 0.01$), $\sigma = 1.41$ ($\pm 0.03$); $t$-model: $\nu = 1.41$ ($\pm 0.21$), $\mu = 0.22$ ($\pm 0.03$); TVERSKY and KAHNEMAN (1992): $\gamma = 0.68$ ($\pm 0.01$). Shaded areas indicate two standard errors in the fitted parameter values. The fit was done by implementing the Levenberg-Marquardt algorithm (LEVENBERG 1944) for non-linear least squares curve fitting.

## 5 Discussion

On 28 February 2020, SUNSTEIN (2020), a behavioural economist, legal scholar, and former United States Administrator of the Office of Information and Regulatory Affairs, diagnosed that people's concern about a potential coronavirus outbreak in the US was attributable to an extreme case of probability weighting. Supposedly, according to Sunstein, people were neglecting the fact that such an event had a low probability. When the piece was published, many commented that it seemed quite reasonable to them to take precautions, and that Sunstein himself may have underestimated both the severity and likelihood of what lay ahead. One month later, the US suffered a major outbreak of coronavirus.

This sad episode illustrates that an inverted S-curve is a neutral indicator of a difference in opinion. It says nothing about who is right and who is wrong.

The term "probability weighting" suggests an obscure mental process, where a DM carries out operations on probabilities. It seems more natural to us to consider a DM modelling events about whose probabilities he is unsure. From this latter point of view, it is easy to think of reasons for a DM's model to differ from a DO's. DMs will often have cause to include additional uncertainty, leading to the frequently observed inverse-S curve.

The model of estimating probabilities from real time series, which we discuss in Sec. 3, has qualitative features that display a degree of universality. Relative errors in the DM's probability estimates are always greater for rarer events. A dislike of the unexpected, which explains the systematic overestimation of low probabilities, is similarly common. "Probability weighting" is purely descriptive and comes with the ill-conceived connotation of DMs suffering from a cognitive error. The phenomenon is better thought of as DMs making wise decisions given the information available to them. Such information is necessarily limited because, for example, DMs are constrained to collect such information in time.

# A  Inverse-S from relative errors in probabilities

For an inverse-S curve to emerge, small probability densities have to be overestimated ($w > p$) and large ones underestimated ($w < p$), as is indeed the case, for example in Fig. 6. Let's connect this statement to one about relative uncertainties. The decision weight is arrived at by adding the probability $p(x)$ to its uncertainty $\varepsilon[p(x)]$ and normalising, as we did in (Eq. 13), *i.e.*

$$w(x) = \frac{p(x) + \varepsilon[p(x)]}{\int_{-\infty}^{\infty} (p(s) + \varepsilon[p(s)])\, ds}\ . \tag{17}$$

This can be expressed as

$$w(x) = p(x) \left( \frac{1 + \frac{\varepsilon[p(x)]}{p(x)}}{\int_{-\infty}^{\infty} p(s)\left\{1 + \frac{\varepsilon[p(s)]}{p(s)}\right\} ds} \right), \tag{18}$$

where $\frac{\varepsilon[p(x)]}{p(x)}$ is the relative error, and the denominator of (Eq. 18) is a normalisation constant. If the relative error is larger for small probabilities than for large probabilities, then small probabilities are enhanced more (the summand $\frac{\varepsilon[p(x)]}{p(x)}$ in the numerator is greater) than large probabilities. The normalisation constant scales down all probabilities equally, and where the enhancement was greater, $w(x)$ ends up above $p(x)$, and where it was lower $w(x)$ ends up below $p(x)$. So, if the relative error is larger for small probabilities, an inverse-S curve emerges.

We can say one more thing about this procedure. If an inverse-S curve exists, then $p(x)$ and $w(x)$ cross somewhere, see Fig. 6. This happens when the relative error attains its expectation value (with respect to the density $p$). Rewriting (Eq. 18) as

$$w(x) = p(x) \left( \frac{1 + \frac{\varepsilon[p]}{p}}{1 + \left\langle \frac{\varepsilon[p]}{p} \right\rangle} \right), \tag{19}$$

we see that $w(x) = p(x)$ when $\frac{\varepsilon[p]}{p} = \left\langle \frac{\varepsilon[p]}{p} \right\rangle$.